\documentclass[12pt,preprint]{aastex}

\newcommand{\myemail}{rmason@noao.edu}

\shorttitle{Hydrocarbon Dust in Seyferts and ULIRGs} 

\shortauthors{Mason et al.}

\begin{document}
\title{Hydrocarbon Dust Absorption in Seyfert Galaxies and
ULIRGs\footnote{Based on observations with the UK Infrared Telescope,
Mauna Kea Observatory, Hawaii.}}

\author{R. E. Mason \altaffilmark{2}} \affil{Institute for Astronomy,
Royal Observatory Edinburgh, Blackford Hill, Edinburgh EH9 3HJ, UK }
\email{\myemail}

\author{G. Wright} 
\affil{UK Astronomy Technology
Centre, Royal Observatory Edinburgh, Blackford Hill, Edinburgh EH9 3HJ, UK}
\email{gsw@roe.ac.uk}

\author{Y. Pendleton}
\affil{NASA/Ames Research Center, Mail Stop 245-3, Moffet Field CA 94035, USA}
\email{ypendleton@mail.arc.nasa.gov}

\and

\author{A. Adamson}
\affil{Joint Astronomy Centre, 660 North A`ohoku Place, Hilo, HI 96720, USA}
\email{a.adamson@jach.hawaii.edu}

\altaffiltext{2}{NOAO Gemini Science Center, 950 N. Cherry Ave., Tucson, AZ 85719, USA}


\begin{abstract}
We present new spectroscopic observations of the 3.4$\mu$m absorption
feature in the Seyfert galaxies, NGC1068 and NGC7674, and the
ultraluminous infrared galaxy, IRAS08572+3915.  A signature of C\sbond
H bonds in aliphatic hydrocarbons, the 3.4$\mu$m feature indicates the
presence of organic material in Galactic and extragalactic dust. Here
we compare the 3.4$\mu$m feature in all the galaxies in which it has
been detected. In several cases, the signal-to-noise ratio and
spectral resolution permit a detailed examination of the feature
profile, something which has rarely been attempted in extragalactic
lines of sight. The 3.4$\mu$m band in these galaxies closely resembles
that seen in the Galactic diffuse ISM and in newly-formed dust in a
protoplanetary nebula. The similarity implies a common carrier for the
carbonaceous component of dust, and one which is resistant to
processing in the interstellar and/or circumnuclear medium. We also
examine the mid-IR spectrum of NGC1068, because absorption bands in
the 5-8$\mu$m region further constrain the chemistry of the 3.4$\mu$m
band carrier.  While weak features like those present in the mid-IR
spectrum of diffuse dust towards the Galactic center would be
undetectable in NGC1068, the strong bands found in the spectra of many
proposed dust analog materials are clearly absent, eliminating certain
candidates and production mechanisms for the carrier. The absence of
strong absorption features at 5-8$\mu$m is also consistent with the
interpretation that the similarity in the 3.4$\mu$m feature in NGC1068
to that in Galactic lines of sight reflects real chemical similarity
in the carbonaceous dust.

\end{abstract}




\section{Introduction}
\label{intro}

Interstellar dust regulates star formation, catalyzes molecule
production and reprocesses UV and optical radiation, thus playing a
critical role both in the physical processes of a galaxy and in our
interpretation of observations of dusty regions. Despite this
importance, however, dust in galaxies other than our own is not yet
well understood.

A useful tracer of the carbonaceous component of dust is the C\sbond H
bond stretch of aliphatic hydrocarbons, observed in absorption at
3.4$\mu$m (2941cm$^{-1}$) throughout the Galactic diffuse ISM. This
broad feature exhibits substructure corresponding to symmetric and
asymmetric stretches of C\sbond H bonds in CH$_{2}$ and CH$_{3}$
groups in aliphatic hydrocarbon chains \citep{S91}, the relative
strengths of which are related to the lengths of the chains
\citep[e.g.][]{Alp}. In addition, chemical groups such as CO, CN, OH,
and aromatic rings attached to the chains can change the force
constants of the C\sbond H bonds, thereby causing shifts in the
positions and intensities of the subpeaks within the band.  The
overall shape of the 3.4$\mu$m feature reflects this and can therefore
be used to probe the chemical makeup of the dust.


Hydrocarbon-containing materials with quite different formation and
processing histories may have similar 3.4$\mu$m band profiles which
reflect the molecular environment of the C\sbond H bonds \citep{P94},
and, as shown by \citet{PandA}, spectroscopy in the 5-9$\mu$m region
is required to differentiate decisively between dust analog materials.
At these wavelengths, features arising from C\sbond H bends, C\sbond C
vibrations and vibrations of many other chemical bonds can also be
present, allowing more complete identification of functional groups
and comparisons between astronomical and laboratory spectra.  However,
while hydrocarbon materials with different histories {\em can} have
similar 3.4$\mu$m profiles, this is not {\em necessarily} the case;
laboratory spectra from a multitude of research groups reveal clear
differences in many of the 3.4$\mu$m feature profiles
\citep{PandA}. Moreover, unlike the 5-9$\mu$m region, the 3.4$\mu$m
feature is accessible from ground-based observatories.
To investigate the extent of these possible variations in the shape of
the 3.4$\mu$m feature, to provide data to complement future mid-IR
spectral observations, and to deepen our understanding of the
carbonaceous dust in the Universe, it will be instructive to compare
spectra of the 3.4$\mu$m feature in different Galactic and
extragalactic environments.

Observations have been made along a number of Galactic lines of sight,
sampling the diffuse ISM towards the Galactic center and a number of
reddened field stars and YSOs, and the newly-formed circumstellar dust
in the protoplanetary nebula, CRL618
\citep{Butch,McFadz,A90,LJ90,S91,P94,S95,Im96,Wh97,Ch98,Chiar00,Chiar02,Ishii,R03}. The
3.4$\mu$m feature has also been studied in the Murchison meteorite
\citep{Murch1,Murch}. Where the quality of the data allows the
comparison, the profile of the feature is very similar between lines
of sight, as well as in some laboratory analog materials
\citep[e.g.][]{P94,Ch98,Ishii,PandA}.

The 3.4$\mu$m band has also been detected in several Seyfert galaxies
and ULIRGs
\citep*{Bridger,Mizutani,boss,Im97,Im00a,Im00b,ID00,Im01,Im02,Im032,MB03,Ris}. In
almost all cases, the observations were aimed at establishing the
power source of each galaxy through the presence or absence of the
3.4$\mu$m band and the 3.3$\mu$m PAH emission feature, rather than to
examine in detail the chemical makeup of the dust.  \citet{boss} and
\citet{Im97} note the similarity of the 3.4$\mu$m feature in the
Seyfert 2 galaxy, NGC1068, with that observed towards the Galactic
center, and \citet{boss} and \citet{P96,P97} also compare the feature
in the ULIRG, IRAS08572+3915, to the Galactic center, again finding a
striking similarity. \citet{Ris} present an L-band spectrum of the
southern nucleus of the Superantennae galaxies (IRAS19254-7245), and
suggest that the 3.4$\mu$m feature profile implies that the
carbonaceous dust in that galaxy contains more electronegative groups
than that in the Galactic diffuse ISM\footnote[3]{Since the submission
of this paper, \citet{Dartois} have also examined the 3.4$\mu$m band,
hydrocarbon/silicate ratio and oxygen content of the carbonaceous dust
in four galaxies. }.


To investigate in more detail the chemical composition of hydrocarbon
dust in galaxies, we have obtained new, higher signal-to-noise ratio
(S/N) spectra of the feature towards the nucleus of NGC1068 and
IRAS08572+3519 (\S\ref{obs}). We also present a new detection of the
feature towards the nucleus of the Seyfert 2 galaxy, NGC7674. In
\S\ref{3.4} we compare the 3.4$\mu$m feature of NGC1068, the galaxy
with the best L-band spectrum, with that observed towards the Galactic
center and in the protoplanetary nebula CRL618, and with all of the
galaxies in which this feature has been detected. Finally, in
\S\ref{MIR} we analyse the 3.4$\mu$m band of NGC1068 in conjunction
with ISO mid-IR spectra of that galaxy, something which has to date
only been possible for the line of sight to the Galactic Center (Chiar
et al. 2000). The implications of this work for the composition,
production and survival of carbonaceous dust are discussed in
\S\ref{discuss}.

\section{The lines of sight}
\label{mygals}

\subsection{The Seyfert galaxies}
\label{1068}

As the galaxy in which \citet{AM85} detected the first hidden broad
line region in polarized light, NGC1068 is probably the most-studied
active galaxy.  At a distance of 16Mpc ($\rm H_{0}=70 km \ s^{-1} \
Mpc^{-1}$), $1^{\prime \prime}$ corresponds to 70pc.  The unified
model of active galactic nuclei (AGN) proposes that our view of the
nucleus of NGC1068, a type 2 Seyfert galaxy, is blocked by an edge-on
dusty torus, and that the broad lines visible only in polarized light
are deflected into our line of sight by a scattering mirror of
electrons or dust. Emission from dust in the central few hundred pc of
NGC1068 can clearly be seen in IR images of that galaxy
\citep[e.g.][]{Tom,Rouan2}. The 0.48\arcsec-wide slit used for the
present observations was centered on the peak of the L-band emission,
and the disk of NGC1068 is quite close to face-on ($i\sim 29^{\circ}$,
see Table~\ref{table}) so the spectrum presented here comes from a
region of the dust surrounding the central engine of this galaxy that
is roughly comparable in size to a Galactic giant molecular cloud
complex.

NGC7674 is also a spiral galaxy with a Seyfert 2 nucleus. Like
NGC1068, the disk of the galaxy has a low inclination, so although
NGC7674 is several times as distant as NGC1068, much of the dust that
we observe is likely to be close to the active nucleus rather than in
the diffuse ISM of the galaxy. The other Seyfert galaxies in which the
3.4$\mu$m feature has been detected have greater inclinations.
In these cases, there may be some contribution from absorption by dust
in the diffuse ISM of the spiral arms of the galaxy.

\subsection{The ULIRGs}
\label{08572}

The deepest 3.4$\mu$m features yet observed in any lines of sight are
found in the ultraluminous infrared galaxies (ULIRGs,
$L_{IR}>10^{12}L_{\odot}$), IRAS08572+3915, IRAS19254-7245 and
UGC5101, along with a weaker feature in IRAS05189-2524
\citep{boss,ID00,Im01,Ris}, leading to the suggestion that the major
power source in these ULIRGs is a deeply obscured AGN. This could mean
that the feature arises in dust close to the active nucleus, as is the
case for some of the Seyfert galaxies. However, unlike the Seyfert
galaxies with a classical spiral morphology, the ULIRGs are often
disturbed, interacting systems, with dust lanes and discs crossing the
nuclear regions \citep{Scov}.
Therefore it is possible that the 3.4$\mu$m band in these galaxies is
caused by a combination of both diffuse medium and circumnuclear dust.

Some relevant properties of the galaxies examined in this paper are
summarized in Table~\ref{table}.


\begin{deluxetable}{cccccc}
\tabletypesize{\scriptsize} \tablecaption{Properties of the galaxies
in which the 3.4$\mu$m feature has been detected \label{table}}
\tablewidth{0pt} 
\tablehead{ \colhead{Galaxy} &
\colhead{Type/morphology \tablenotemark{a}} &
\colhead{Redshift \tablenotemark{a}} & 
\colhead{Inclination \tablenotemark{b}} &
\colhead{N$_{H}$  \tablenotemark{c}} &
\colhead{References}\\ 
\colhead{} &
\colhead{} &
\colhead{} & 
\colhead{(deg)} &
\colhead{(10$^{20}$ cm$^{-2}$)} &
\colhead{} }

\startdata
NGC1068 & Sy2/SAb & 0.0038 & 29$^{\circ}$ & $\ge 10^{5}$ & 1,2,3,4\\
IRAS04385-0828 & Sy2/S0 & 0.0151 & $>$60$^{\circ}$ & ...  & 11\\
IRAS05189-2524 & ULIRG, Sy2/-- & 0.0426 & ... & 490$^{+10}_{-16}$ & 1,2,5\\
IRAS08572+3915 & ULIRG, Sy2/-- & 0.0584 & ...  & $\ge 10^{5}$ & 1,2,5 \\
UGC5101 & ULIRG, Sy1.5/-- & 0.0394 & ... & $1.4 \times 10^{4}$& 6\\
NGC3094 & Sy2/SBa & 0.0080 & 48$^{\circ}$ & ... & 7 \\
Mrk463 & Sy2/-- & 0.0510  & 55$^{\circ}$ & 1600$^{+800}_{-800}$& 9 \\
IRAS19254-7245 & ULIRG, Sy2/-- & 0.0617 & ...  & $\ge 10^{5}$ & 10\\
NGC7172 & Sy2/Sa & 0.0087 & 61$^{\circ}$  & 861$^{+79}_{-33}$  & 7 \\
NGC7479 & Sy2/SBc &  0.0079 & 51$^{\circ}$  & ...  & 7 \\
NGC7674 & Sy2/SAbc & 0.0289 & 23$^{\circ}$  & $\ge 10^{5}$ & 1,11\\

\enddata


\tablenotetext{a}{From NED} 

\tablenotetext{b}{Spiral galaxies only. From \citet{Whittle}, except:
NGC1068, NGC7476 --- \citet{Garcia}; NGC7172 --- \citet{Nagar};
NGC3094 --- \citet{RC3}; IRAS04385 --- \citet{Kinney}. }

\tablenotetext{c}{From \citet{B99}, except: IRAS08572, IRAS19254 ---
\citet{Ris00}; UGC5101 --- \citet{Imetal}}



\tablerefs{
[1]~This work
[2]~\citet{boss}
[3]~\citet{Im97}
[4]~\citet{MB03}
[5]~\citet{ID00}
[6]~\citet{Im01}
[7]~\citet{Im00b}
[8]~\citet{Im00a}
[9]~\citet{Im02}
[10]~\citet{Ris}
[11]~\citet{Im032}
}
\end{deluxetable}


\section{Observations and data reduction}
\label{obs}

Spectra of NGC7674 and IRAS08572 were obtained using CGS4 on the 3.8m
United Kingdom Infrared Telescope (UKIRT) on Mauna Kea, in October
2000. The observations of IRAS08572 were completed in March 2001, as
part of the UKIRT Service Programme. For NGC7674, the 40 l/mm grating
was used with a 2-pixel (1.2'') slit, giving a resolving power of
$\sim 700$. IRAS08572 was observed with the 150 l/mm grating and the
2-pixel slit, resulting in R$\sim 2600$. NGC1068 was observed in
November 2002 with UIST, UKIRT's imager/spectrometer. UIST's short-L
grism and 4-pixel (0.48'') slit were used, for R$\sim 650$.

The data were reduced using UKIRT's data reduction pipelines, CGS4DR
and ORACDR, and the Starlink FIGARO package. After bad pixel masking,
bias subtraction, and flatfielding, subtracted sky and object pairs
were coadded into groups. Residual sky emission, which was generally
very low, was removed by fitting a polynomial in the spatial direction
and the spectra were then optimally extracted. Sky lines were
cancelled by division by a standard star. For IRAS08572 and NGC7674,
wavelength calibration was achieved using observations of an argon arc
lamp, but for the UIST observations of NGC1068, the arc lamp was
observed through the H-band filter in second order, which induces a
wavelength shift in the lines. Rather than correcting for this shift,
HI lines in the standard star and the telluric CH$_{4}$ line were used
for wavelength calibration.

The spectra of IRAS08572 and NGC7674 were obtained using more than one
grating setting. The separate spectra were merged by wavelength, being
multiplied by a constant where necessary to achieve a good match in
flux between the sections (these adjustments were generally $<$ 10\%)
The wavelength calibration should ensure a good match in wavelength,
but this was checked using telluric lines where possible. The spectra
were flux-calibrated using the V- or K-band magnitudes of the telluric
standard stars and appropriate V-K, K-L and K-L' colors. The results
are generally consistent with published measurements, but as the
standard stars were selected primarily to ensure good cancellation of
atmospheric lines rather than as photometric standards, the flux
calibration should be considered approximate.

Data for NGC3094, NGC7172, NGC7479 and Mrk463, which were also
acquired using CGS4 on UKIRT, were obtained from the UKIRT archive and
reduced in the same way as the galaxies described above. These data
have previously been published by \citet{Im00a,Im00b,Im02}; the
observational details are described therein. The L-band spectrum of
IRAS19254 was aquired with the Infrared Spectrometer and Array Camera,
ISAAC, on the Very Large Telescope (VLT), and has been published by
\citet{Ris}. We obtained the data from the VLT Science Archive
Database. Initially, the eclipse-isaac software package \citep{Dev}
was used to flatfield, sky-subtract, wavelength-calibrate and combine
the individual frames. Removal of residual sky emission, optimal
extraction, division by a standard star and flux calibration were
performed using FIGARO routines.

Spectra of UGC5101 and IRAS04385, observed with NSFCAM and SpeX at
NASA's IRTF, were kindly provided by M. Imanishi. The observations and
reduction of the data are described in Imanishi et al.~(2001) and
\citet{Im032}. The spectrum of IRAS05189 is from \citet{boss}.

\subsection{Optical depth spectra}

\begin{figure}
\epsscale{0.9}
\plotone{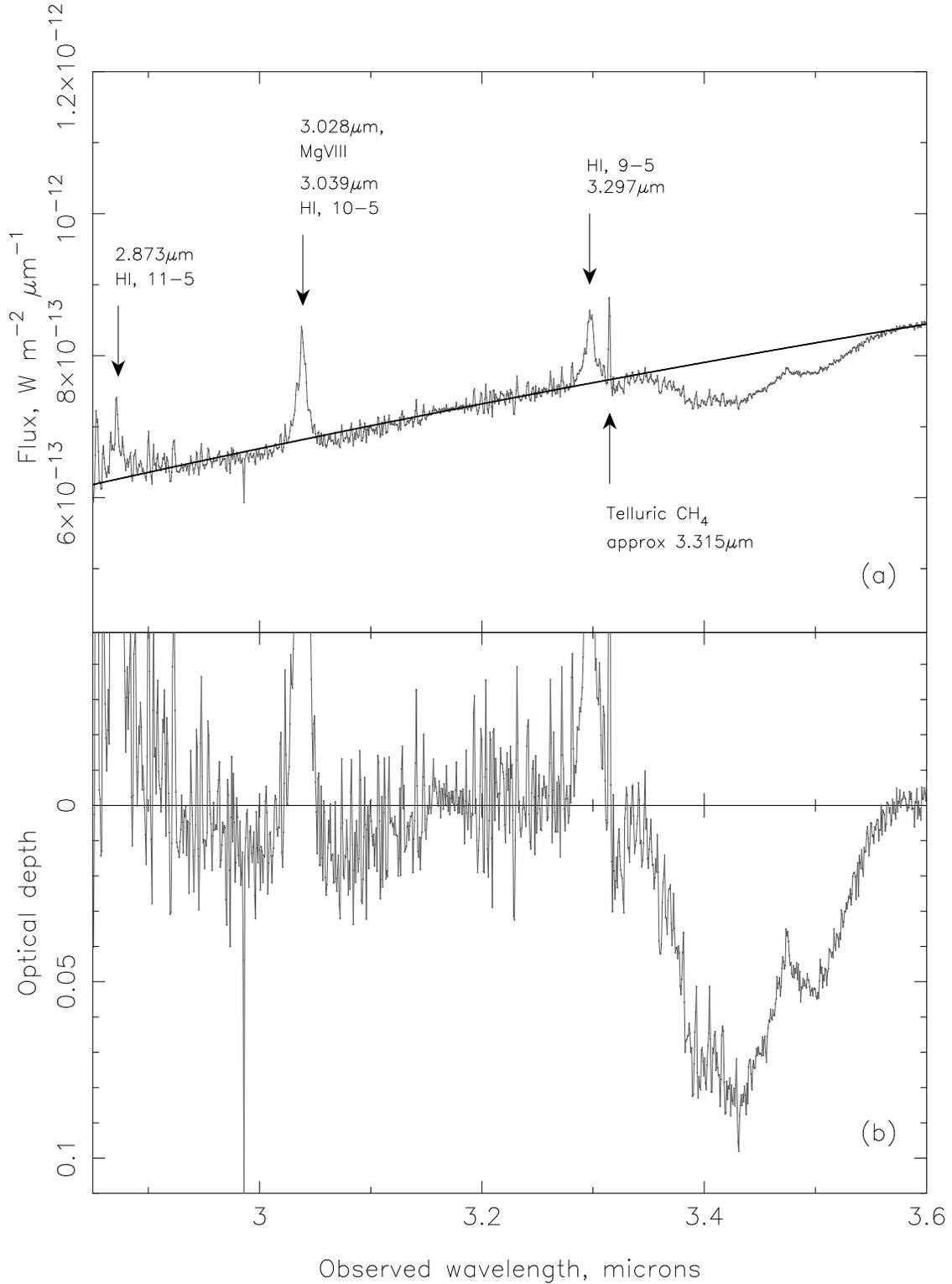} 
\caption{(a) UKIRT/UIST flux spectrum of NGC1068 (R~600), showing the
continuum used to derive the optical depth spectrum. Features arising
in the atmosphere and the telluric standard star are indicated, along
with the 3.028$\mu$m (3.039$\mu$m at z=0.0038) MgVIII line in
NGC1068. (b) Optical depth spectrum of NGC1068.}
\label{fig:1068}
\end{figure}

\begin{figure}
\epsscale{0.9}
\plotone{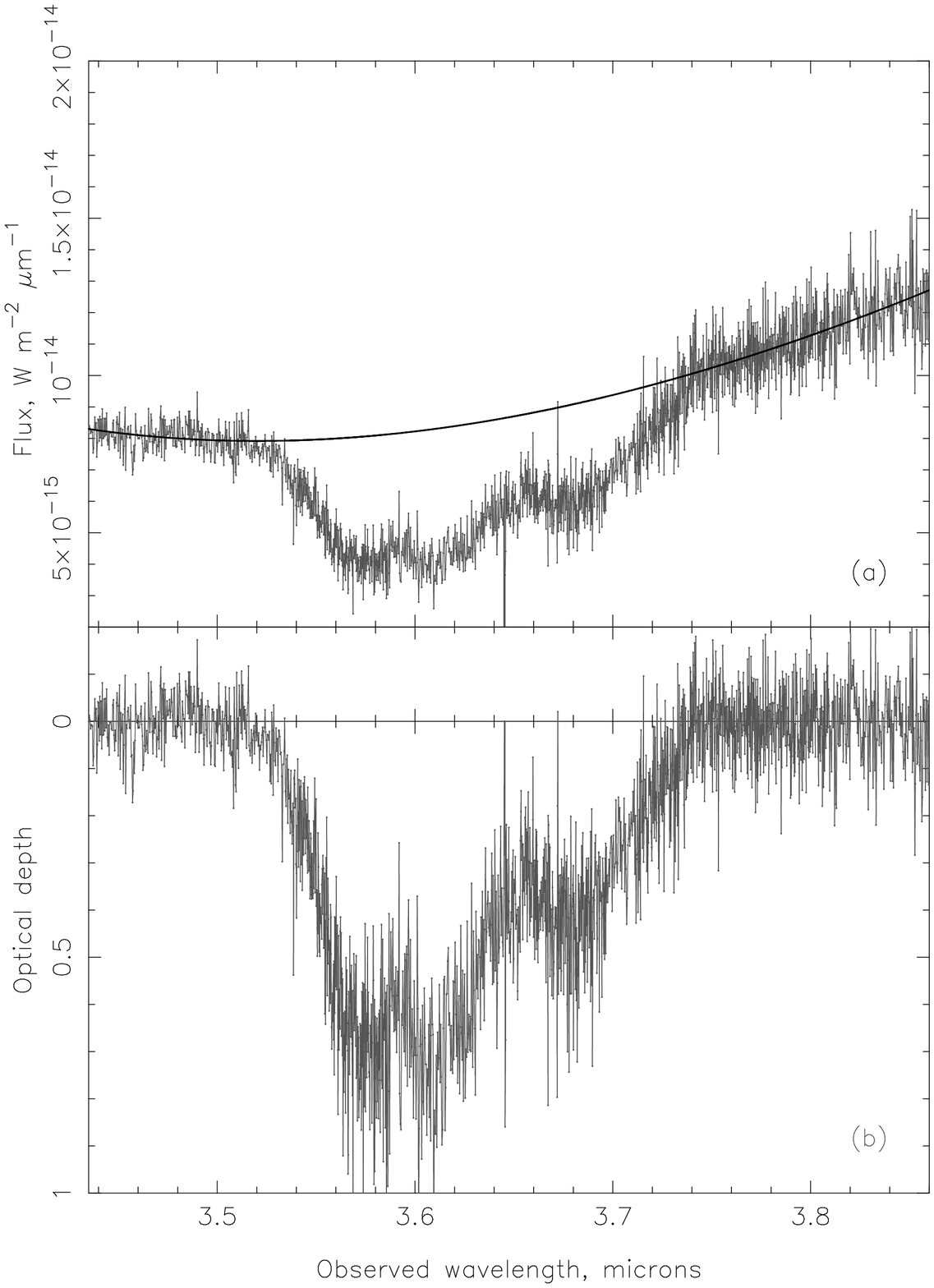}
\caption{(a) UKIRT/CGS4 flux spectrum of IRAS08572 (R~2600), showing the
continuum used to derive the optical depth spectrum. (b) Optical depth
spectrum of IRAS08572.}
\label{fig:08572}
\end{figure}

\begin{figure}
\epsscale{0.9}
\plotone{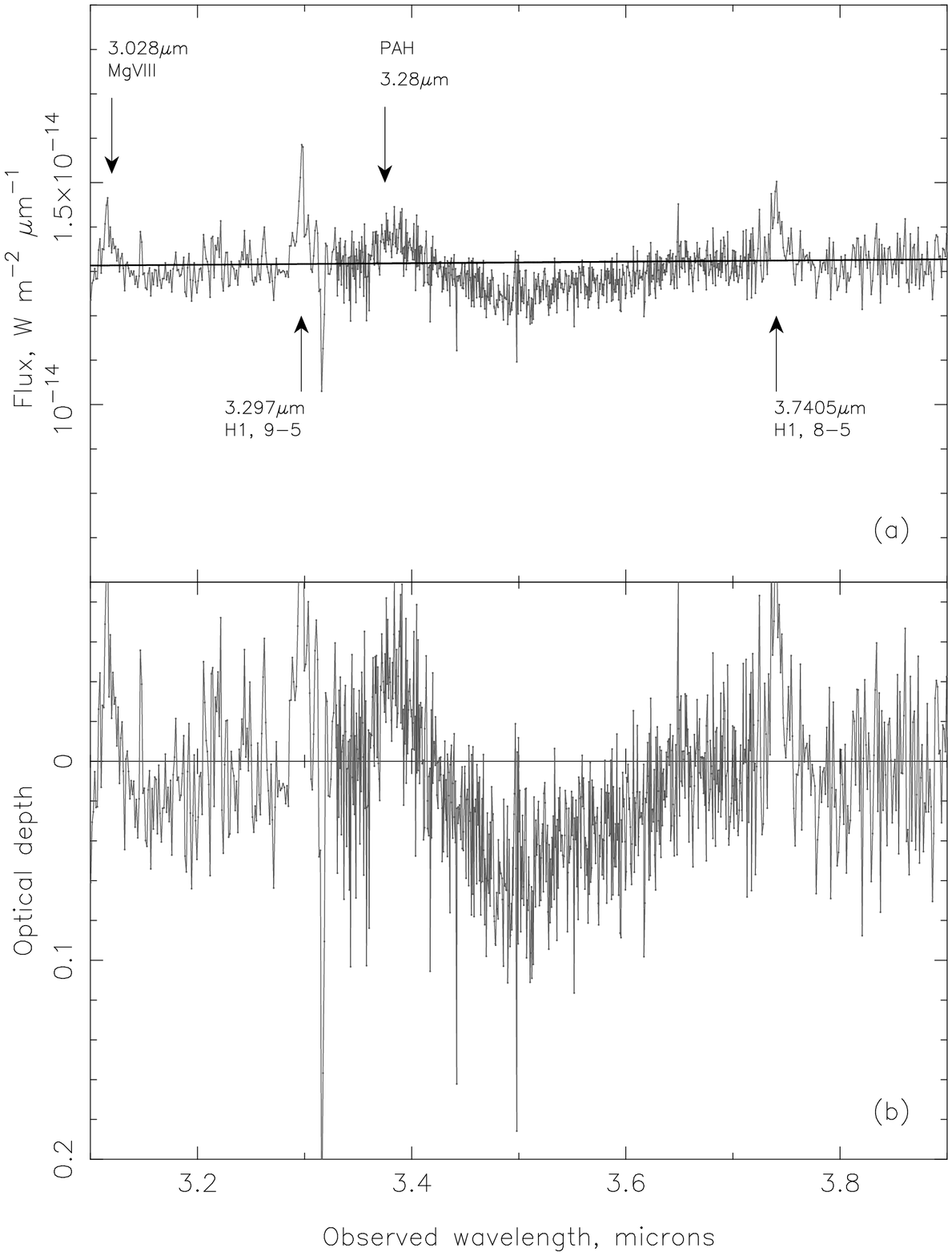}
\caption{(a) UKIRT/CGS4 flux spectrum of NGC7674 (R~700), showing the
continuum used to derive the optical depth spectrum. The 3.28$\mu$m
PAH band and 3.028$\mu$m MgVIII line in NGC7674 (3.38 and 3.116$\mu$m
at z=0.029, respectively) and features arising in the A-type telluric
standard star are marked. (b) Optical depth spectrum of NGC7674.}
\label{fig:NGC7674}
\end{figure}

Flux and optical depth spectra from the new data, together with the
continua used to derive the optical depths, are presented in
Figures~\ref{fig:1068}, \ref{fig:08572} and \ref{fig:NGC7674}. These
continua result from fitting low-order polynomial curves to the
spectra outside the rest-frame 3.30-3.55$\mu$m region.


Several groups have measured the optical depth of the 3.4$\mu$m
feature in NGC1068. \citet{Im97} calculate
$\tau_{3.42}=0.126\pm0.014$, slightly larger than the
$\tau_{3.42}=0.1$ derived by \citet{boss}. We measure
$\tau_{3.42}\approx 0.09$. More recently, \citet{MB03} obtained a
spectrum of NGC1068 in the 3.0-3.6$\mu$m range, with
$\tau_{3.4}=0.14$.
The differences in optical depth values between these studies probably
result from differences in aperture sizes and in the continua used to
derive $\tau_{3.4}$.

L-band spectra of IRAS08572 have been obtained by \citet{boss} and
\citet{ID00}, at $\lambda/\Delta\lambda\sim380$ in the latter
case. Their spectra both have $\tau_{3.42}\sim0.7$, as does the
present spectrum. For NGC7674 we measure $\tau_{3.4}\sim0.07$, and
also detect weak PAH emission at 3.3$\mu$m. This is the first time
that the 3.4$\mu$m feature has been reported in this galaxy; the
presence of the feature is not clear in the lower-resolution spectrum
reported by \citet{Im032}, who find $\tau_{3.4}<0.2$. In none of these
galaxies is there any clear indication of the 3.1$\mu$m O\sbond H
stretch characteristic of the ices observed in Galactic dense
molecular clouds.

\section{Comparison of the 3.4$\mu$m feature in Seyferts, ULIRGs and other dusty environments} 
\label{3.4}

\begin{figure}
\epsscale{0.8}
\plotone{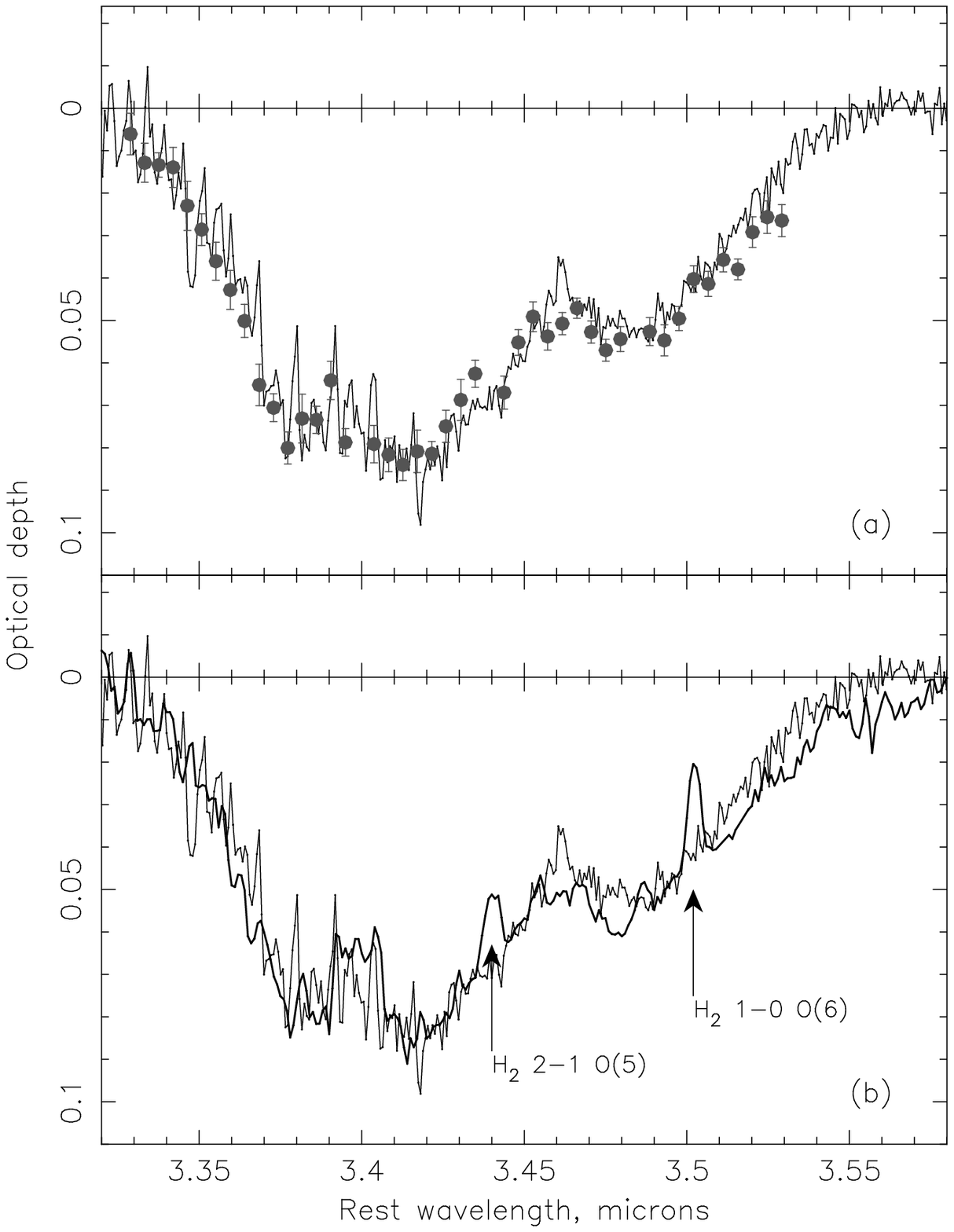}
\caption{The 3.4$\mu$m feature in NGC1068 compared with (a) that in
the diffuse ISM towards the Galactic center source, IRS6E
\citep[][points]{P94} and (b) newly-formed dust in the protoplanetary
nebula, CRL618 \citep[][dark line]{Ch98}. The spectra of IRS6E and
CRL618 have been normalized to NGC1068 at $\tau_{3.42}$, and lines of
collisionally-excited H$_{2}$ indicated on the spectrum of CRL618. The
errors in the spectrum of NGC1068 are approximately the same as the
observed point-to-point scatter, except in the regions of poor
atmospheric cancellation shortwards of approx. 3.42$\mu$m, where
systematic cancellation effects dominate. }
\label{fig:compare1}
\end{figure}

\begin{figure}
\epsscale{0.9}
\plotone{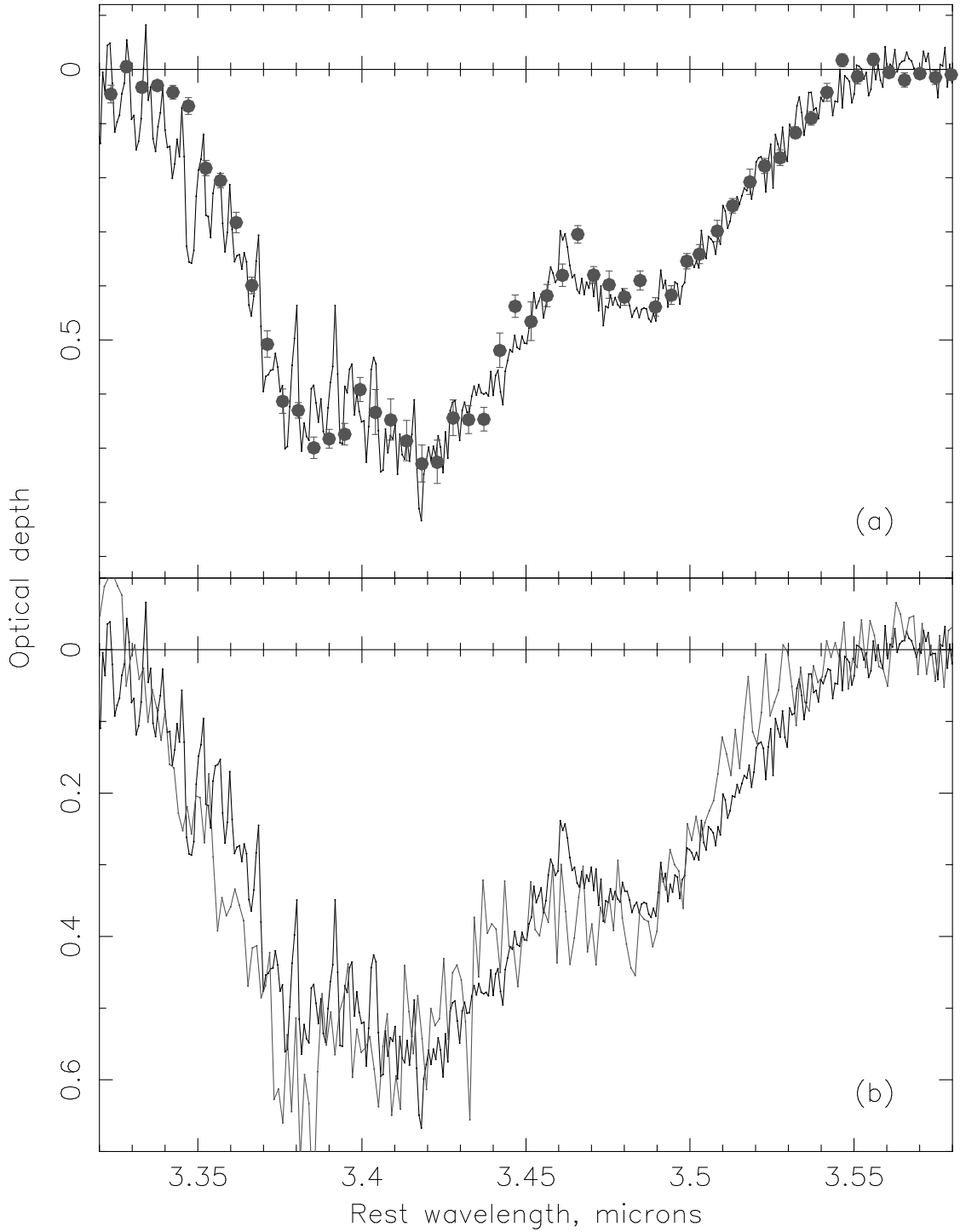}
\caption{The 3.4$\mu$m feature in NGC1068 compared with that in (a)
IRAS08572 and (b) IRAS19254 \citep{Ris}. In each case, NGC1068 is the
solid black line and has been multiplied by a constant in order to
have the same optical depth as the comparison galaxy around
3.42$\mu$m.}
\label{fig:compare2}
\end{figure}

\begin{figure}
\epsscale{0.85}
\plotone{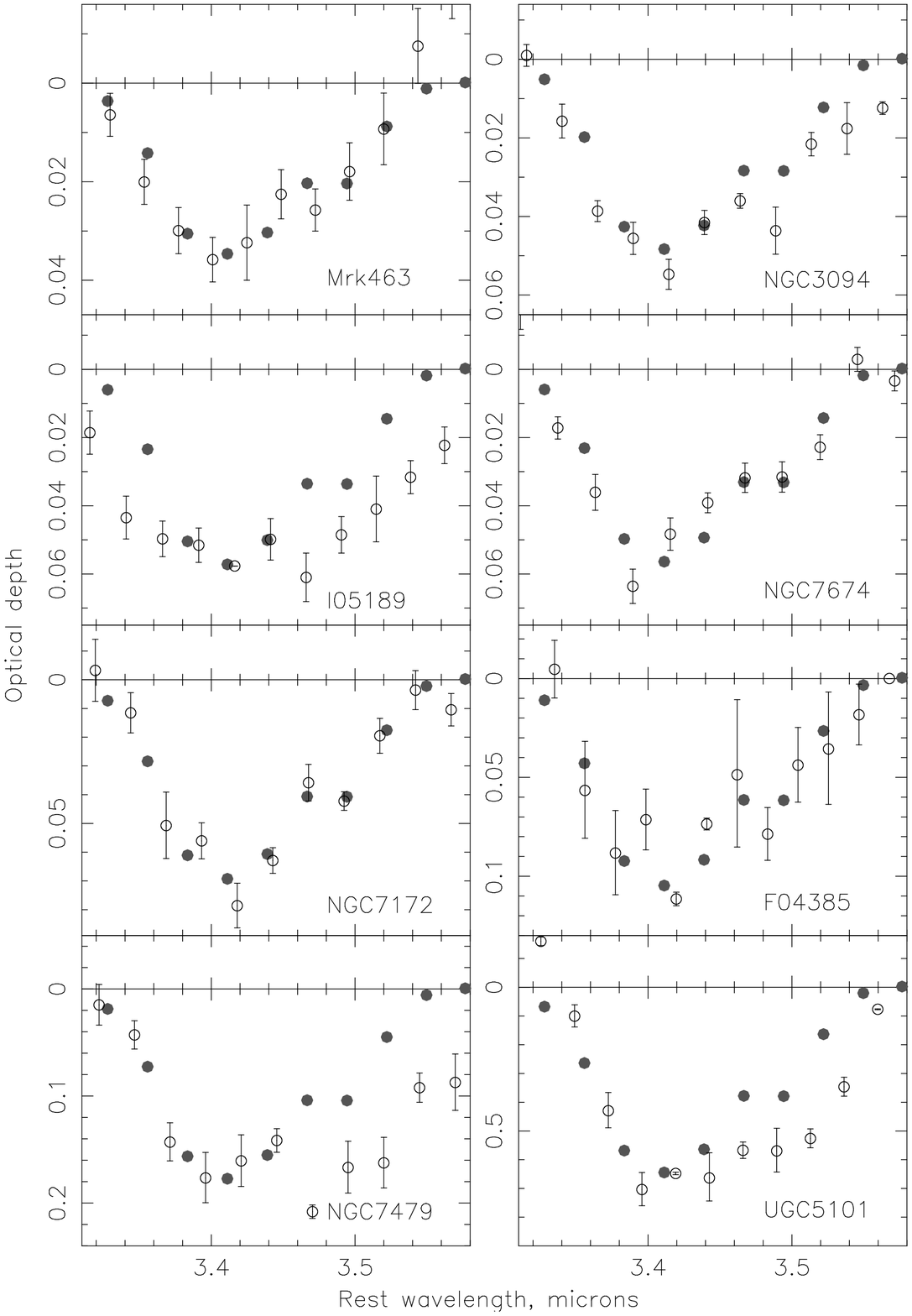}
\caption{The 3.4$\mu$m feature in NGC1068 (filled circles) compared
with that in several other Seyfert 2 galaxies and ULIRGs, displayed in
order of increasing $\tau_{3.4}$. The spectra are shown binned to a
common resolution of $\sim$120, error bars for NGC1068 are smaller
than the data points. References for the data shown in this figure can
be found in Table~\ref{table}.}
\label{fig:compare3}
\end{figure}

Fig.~\ref{fig:compare1} compares the 3.4$\mu$m feature in NGC1068 with
the 3.4$\mu$m feature observed in the diffuse ISM towards the Galactic
center source, IRS6E, and in the protoplanetary nebula, CRL618. In
Figures~\ref{fig:compare2} and \ref{fig:compare3}, the 3.4$\mu$m
feature in NGC1068 is compared with that of the ten other galaxies in
which the feature has been clearly detected to date\footnote[4]{The
3.4$\mu$m absorption has also been seen in a twelfth galaxy, NGC5506
\citep{Im00a}. However, the weakness of the feature in that galaxy
($\tau_{3.4} \le 0.025$) means that its detection, depth and profile
are critically dependent on the choice of baseline against which
$\tau_{3.4}$ is measured. Atmospheric absorption lines at the short
end of the spectrum, which do not cancel out well in this dataset,
make it impossible to unambiguously determine this baseline, so we do
not consider NGC5506 any further.}.

Probably the most striking aspect of Fig.~\ref{fig:compare1} is the
close resemblance that the 3.4$\mu$m feature in NGC1068 bears to that
observed in the Galactic lines of sight. The 3.4$\mu$m absorption in
this Seyfert 2 nucleus clearly exhibits the same subpeaks that are
observed in the feature in the Galactic diffuse ISM and the
protoplanetary nebula: the 3.38$\mu$m CH$_{3}$ asymmetric
stretch, 3.42$\mu$m CH$_{2}$ asymmetric stretch and
3.48$\mu$m CH$_{3}$ symmetric stretch. As in the other spectra,
the 3.50$\mu$m CH$_{2}$ symmetric stretch band is not
clearly present, an observation interpreted by \citet{S91} as the
possible effect of electron-withdrawing groups attached to the
hydrocarbon chains which give rise to the feature (see
\S\ref{discuss.1}). Not only are the peak wavelengths good matches,
but the relative strengths of the subfeatures are also very similar
between NGC1068 and the other objects. This indicates that the
relative numbers of CH$_{3}$ and CH$_{2}$ groups,
and therefore the average lengths of the hydrocarbon chains, are very
similar in all the lines of sight. The new NGC1068 data confirm in
more detail the findings of \citet{boss} and \citet{Im97}.

Besides the Galactic diffuse ISM and CRL618, the 3.4$\mu$m feature in
NGC1068 is very much like the feature in IRAS08572 and IRAS19254
(Fig.~\ref{fig:compare2}). Hydrocarbons in IRAS08572, in particular,
produce a 3.4$\mu$m band which is virtually identical to that caused
by hydrocarbons in the nuclear region of NGC1068. Some small
differences appear in the spectrum of IRAS19254. The feature in this
ULIRG exhibits the same subfeatures and general shape, but, in the
binned spectrum presented by \citet{Ris}, the peak around 3.48$\mu$m
is slightly stronger than that in NGC1068 and the Galactic center. In
our new reduction of these data this difference is less noticeable,
but the 3.38$\mu$m CH$_{3}$ asymmetric stretch is a little more
prominent than in the other lines of sight. Overall, though, the
feature bears quite a close resemblance to that in all of the other
lines of sight examined so far.

Although at much lower S/N, about half of the remaining galaxies show
a similar pattern of a peak near 3.42$\mu$m accompanied by a shoulder
around 3.48$\mu$m, in roughly the same proportions as in NGC1068 and
the Galactic sightlines (Fig.~\ref{fig:compare3}). Interestingly, the
remainder of the sample hints at differences. In particular, the ratio
$\tau_{3.42}/\tau_{3.48}$ may be lower in UGC5101, NGC7479 and
IRAS05189 than in NGC1068. However, in the absence of higher-S/N data
on the 3.4$\mu$m feature in these galaxies, further speculation seems
premature.

\section{The nuclear mid-IR spectrum of \\ NGC1068}
\label{MIR}

As dust absorption features in the mid-IR have been shown to be a
powerful tool for discriminating between models for dust composition
and evolution \citep{PandA}, it would be instructive to compare
5-9$\mu$m spectra of these galaxies with each other and with the
laboratory data. To date, such a comparison has been attempted for
only one line of sight, towards Sgr A* in the Galactic center
\citep{Chiar00,PandA}. The mid-IR spectrum of the diffuse interstellar
dust in that line of sight is remarkably sparse, the only features
being the asymmetric and symmetric bending modes of C\sbond H bonds,
at 6.85 and 7.25$\mu$m respectively. In this regard the spectrum of
Sgr A* is similar to the spectra of plasma-processed carbon materials
(representing dust formed in stellar outflows), but very different
from the more complex spectra of processed ice residues and the
Murchison meteorite spectrum of \citet{Murch}, which display
additional features due to functional groups such as NH, CO and
OH. All of these classes of material can exhibit a 3.4$\mu$m feature
which is a close match to the diffuse ISM data, yet they vary greatly
at longer wavelengths.

Of the galaxies with the best-quality L-band spectra, no mid-IR data
yet exist for IRAS08572. PAH emission bands complicate the
interpretation of the MIR spectrum of IRAS19254 \citep{Charm}. An
ISOCAM CVF 5-16$\mu$m spectrum of the central 700pc
($\approx$10\arcsec, $\lambda/\Delta\lambda\sim30-40$) of NGC1068 has
been published by \citet{LeF}, while at higher spectral resolution
($\lambda/\Delta\lambda\sim1200$), ISO-SWS has covered the full
2-45$\mu$m range \citep{Lutz,Sturm}. The aperture through which the
ISO-SWS spectrum was observed covers 14\arcsec$\times$20\arcsec~at
wavelengths shorter than 12$\mu$m, large enough to include some flux
from the kiloparsec-scale region of star formation in
NGC1068. However, the weakness of the PAH emission bands in this
spectrum and their similarity to those in the smaller-aperture
ISOCAM-CVF spectrum suggests that the ISO-SWS data are dominated by
the region of dust close to the nucleus towards which we measured the
3.4$\mu$m feature. We therefore use this spectrum to assess whether
the similarity of the 3.4$\mu$m feature in NGC1068 to that observed in
the Galactic diffuse ISM is accompanied by a mid-IR spectrum also
resembling that of the Galactic diffuse ISM.

\begin{figure}
\plotone{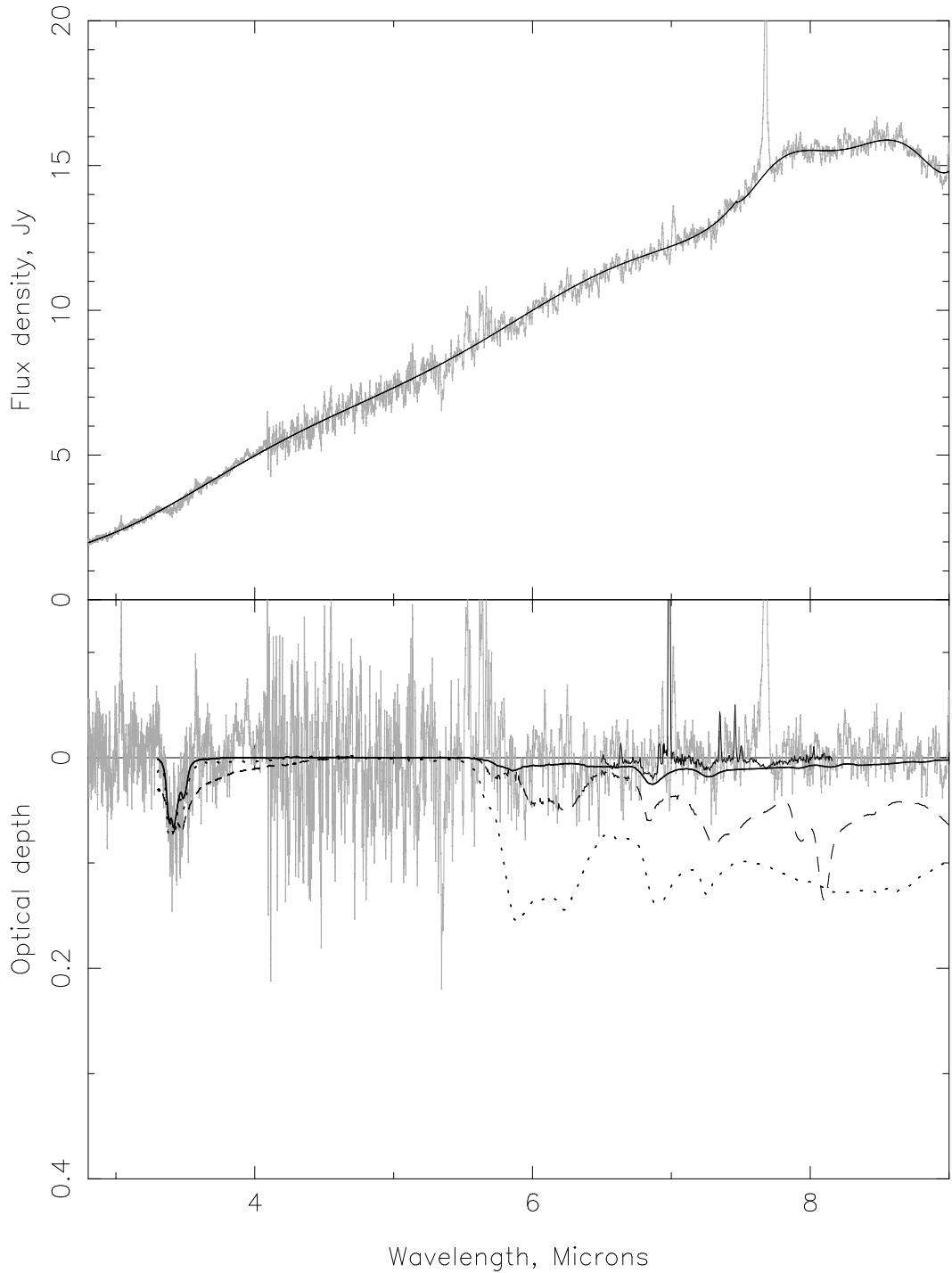}
\caption{(a) ISO-SWS spectrum of NGC1068 \citep{Lutz,Sturm} showing
the baseline used to derive an optical depth spectrum. (b) Mid-IR
optical depth spectrum of NGC1068 compared with other lines of sight
and dust analog materials. In increasing order of optical depth at
7$\mu$m: Sgr A* \citep{Chiar00}; HAC \citep{Men99}; processed ice
residue \citep{A88}; Murchison meteorite \citep{Murch}. }
\label{fig:MIR}
\end{figure}

To establish the presence or absence of mid-IR absorption features in
NGC1068, the spectra were fitted with polynomial curves, with
wavelengths $<7.5\mu$m and $>7.5\mu$m treated separately
(Fig.~\ref{fig:MIR}). The broad emission structure between
approximately 7.5 and 9$\mu$m is a blend of two weak PAH bands
\citep{Sturm} and any hydrocarbon absorption features would be
expected to be seen superimposed on this, so no attempt was made to
find the underlying continuum in this region. As with the 3.4$\mu$m
feature, the fitted curves were used as baselines against which the
optical depth of any absorption features in the spectrum could be
measured. We measure $\tau_{3.4}\approx 0.07$ for the ISO-SWS spectrum
of NGC1068. The smaller value of $\tau_{3.4}$ in this spectrum than in
Fig.~\ref{fig:1068} is likely to be due to the large ISO beam
encompassing some less deeply obscured material than the smaller UIST
aperture.

Also shown in Fig.~\ref{fig:MIR} is a comparison of the NGC1068 ISO
spectra with optical depth spectra of the line of sight towards Sgr~A*
in the Galactic center, the Murchison meteorite, and two laboratory
materials which have been proposed as analogs of the hydrocarbon
component of dust \citep*{A88,Murch,Men99,Chiar00}. These particular
laboratory substances, a hydrogenated amorphous carbon (HAC) and the
residue from a UV-irradiated ice, were chosen because, while they both
have absorption at 3.4$\mu$m, they have very different 5-9$\mu$m
spectra \citep{PandA}. The HAC sample produces some absorption below
3$\mu$m and between the 3.4$\mu$m feature and the 5-9$\mu$m features,
but such featureless absorption would be indistinguishable from the
continuum level in the NGC1068 spectrum. This absorption was therefore
subtracted from the HAC spectrum.  All the comparison spectra have
been multiplied by a constant in order to be normalized to
$\tau_{3.4}=0.07$.

The detailed similarity of NGC1068 and Sgr~A* is difficult to
ascertain. It is apparent from Fig.~\ref{fig:MIR} that if dust
resembling that in the Galactic diffuse ISM is present in the nucleus
of NGC1068, we should not expect to be able to detect and compare the
6.85 and 7.25$\mu$m C\sbond H bends which accompany the 3.4$\mu$m
band. The weakness of these features in the Galactic center spectrum,
approximately 0.24 and 0.14 times the depth of the 3.4$\mu$m feature
in that line of sight \citep{Chiar00}, would make them undetectable in
the spectra of NGC1068. The same is true for the laboratory spectrum
of the HAC \citep{Men99}. This leaves open the possibility that weak
bands exist in NGC1068 which are not observed in the Galactic
center. On the other hand, Fig.~\ref{fig:MIR} suggests that if the
strong bands arising in the spectra of the ice residue \citep{A88} and
the meteorite \citep{Murch} were present in the NGC1068 data, they
would be clearly visible. As with the 3.4$\mu$m feature, the dust in
NGC1068 could have had a very different 5-9$\mu$m spectrum from that
of Galactic dust, but large differences are not observed.

\section{Discussion}
\label{discuss}

\subsection{Interpreting small changes in the profile of the 3.4$\mu$m feature}
\label{discuss.1}
 
In \S\ref{3.4}, it was noted that the 3.4$\mu$m feature in IRAS19254,
while overall a good match to that observed in other lines of sight,
may show some small differences in the relative depths of the subpeaks
within the feature. \citet{Ris} suggest that the higher depth of the
3.48$\mu$m subfeature in their spectrum of IRAS19254 may be attributed
to increased numbers of electronegative groups in the grains.
In pure saturated aliphatic hydrocarbons (compounds with the general
formula C$_{n}$H$_{2n+2}$, e.g. propane, butane), four subpeaks are
visible within the 3.4$\mu$m feature, rather than the three which can
be seen in the astronomical spectra. The four peaks are caused by
symmetric and asymmetric stretches of CH$_{2}$ and CH$_{3}$ groups. As
the symmetric stretch features lie closer together than the asymmetric
stretch bands, shifts in band frequencies caused by nearby
electron-withdrawing groups are more likely to cause blending of the
symmetric stretch features. This effect could account for the apparent
disappearance of the 3.50$\mu$m CH$_{2}$ symmetric stretch band in the
astronomical spectra, which may in fact simply have been shifted to a
position coincident with the 3.48$\mu$m CH$_{3}$ subfeature
\citep{S91}.

Therefore it is plausible that an increase in the number of
electronegative groups could shift the CH$_{2}$ symmetric stretch band
further under the CH$_{3}$ symmetric stretch band, marginally
enhancing the depth of the CH$_{3}$ subfeature at
3.48$\mu$m. Alternatively, given the slight increase in the strength
of the CH$_{3}$ asymmetric stretch band at 3.38$\mu$m in IRAS19254, we
could be observing a decrease in the typical length of the hydrocarbon
chains in the dust. Or, conversely, an increase in the length of the
chains could be supressing the symmetric stretch features with respect
to the asymmetric stretching bands \citep{Cerni}. In fact, while
overall trends in the shape of this feature are governed by factors
such as chain length and the presence of elements other than C and H
in the chains, these factors act together in a very complex manner to
influence the details of the feature profile. Although variations in
the shape of the 3.4$\mu$m feature undoubtedly reflect the chemical
composition of the material which gives rise to it, it is difficult to
ascribe small changes in the 3.4$\mu$m feature in complex mixtures of
hydrocarbons to particular, specific chemical changes with much
confidence.

\subsection{The chemical makeup of the carbonaceous dust in other galaxies}
\label{discuss.2}

It is clear that, certainly in the galaxies with the best data ---
NGC1068, IRAS08572 and IRAS19254 --- the 3.4$\mu$m feature varies only
slightly from galaxy to galaxy and is very similar to that observed in
Galactic lines of sight. In addition, about half of the other galaxies
with lower-S/N detections of the 3.4$\mu$m band have features with
profiles consistent with that of NGC1068.

One explanation for the similarity of the 3.4$\mu$m feature in several
different galaxies may be that the spectra of different
hydrocarbon-containing substances are simply averaging out to the
observed spectrum. This may well be the case for small variations in
the band profile, but it is less plausible that large, systematic
differences in the 3.4$\mu$m band in different environments could be
masked in this way. If, for instance, dust in IRAS08572 generally
contained a much higher fraction of longer hydrocarbon chains than
dust in NGC1068, this would be expected to enhance the subpeaks
corresponding to CH$_{2}$ groups. That some of the galaxies in this
sample do appear to show differences in the band profile, as do
complex laboratory mixtures of hydrocarbons, supports this view. 
It appears that, whatever other differences there may be in the
dust in other galaxies, the hydrocarbon fraction bears a close
resemblance to that which we observe in Galactic lines of sight.

\begin{figure}
\plotone{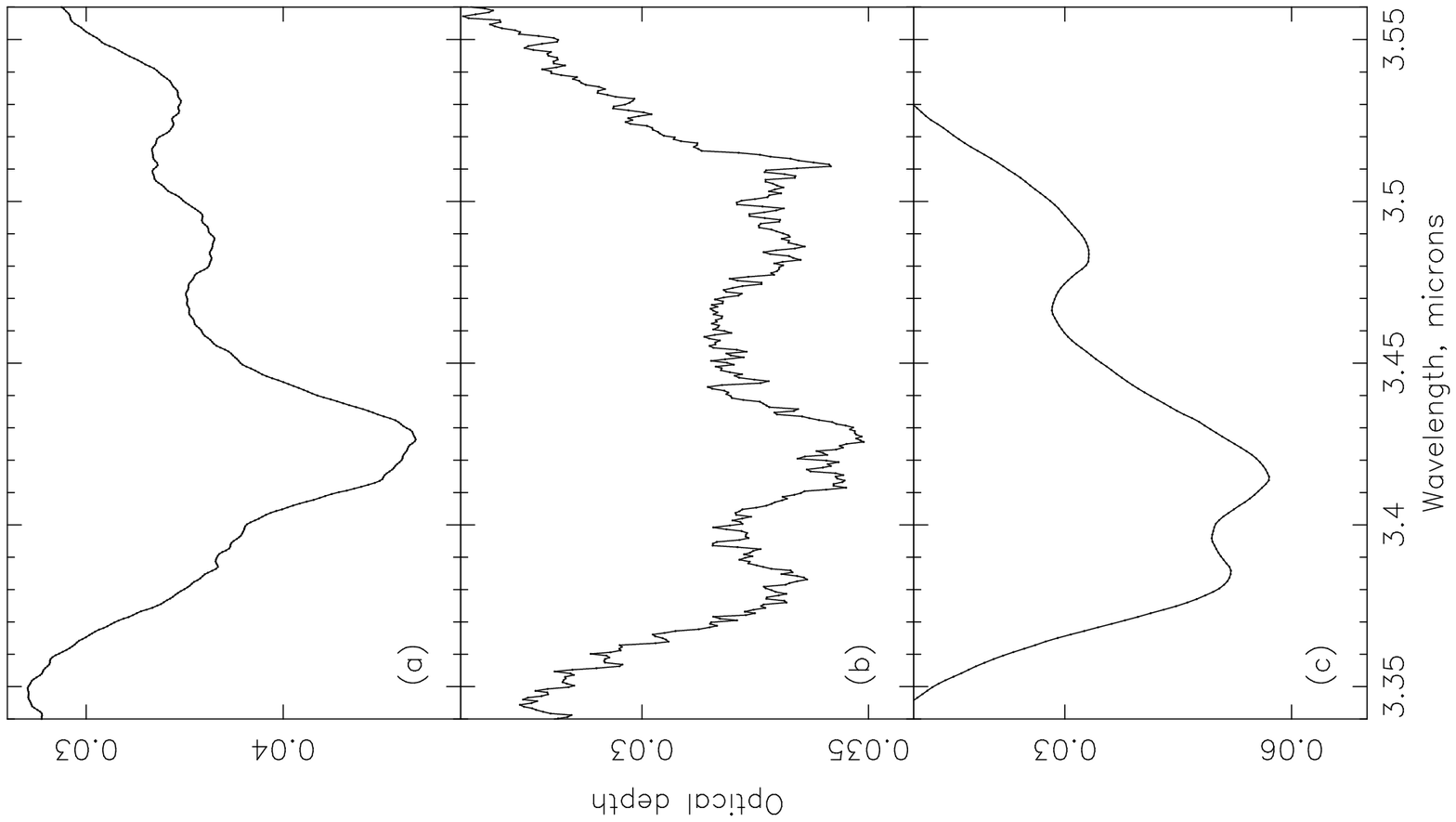}
\caption{The 3.4$\mu$m feature in three different
hydrocarbon-containing substances. {\em (a)} A quenched carbonaceous
condensate \citep{S87} ;{\em (b)} A UV-irradiated ice \citep{A88}; {\em
(c)} An interstellar HAC analog \citep{Furton}}
\label{fig:labspecs}
\end{figure}

While the 3.4$\mu$m feature relates directly only to the saturated
hydrocarbon chains, the similarity of the band in several different
lines of sight may actually reflect overall compositional similarity
in the dust. Based on fits to the 3.4$\mu$m feature, a number of
analog materials produced by different mechanisms have been suggested
to represent the carbonaceous component of dust. \citet{P94} show that
certain processed carbon-rich ice residues can produce about as good a
match to the interstellar 3.4$\mu$m feature as those produced from
carbon plasmas, for example. Only inconsistencies with observations in
the 5-9$\mu$m region rule out processed ices as good analogs of
interstellar dust \citep{PandA}. This means that the 3.4$\mu$m
absorption alone is not the ideal tool with which to probe dust
composition and evolution. On the other hand, there also exist plenty
of dust candidate materials with 3.4$\mu$m bands which clearly differ
from each other and from the astronomical spectra. This is illustrated
in Fig.~\ref{fig:labspecs}, which compares the 3.4$\mu$m band in three
different hydrocarbon-containing substances. Band profiles as varied
as these would be easily distinguishable in spectra with the S/N and
resolution as those of at least three of the galaxies for which L-band
data exist.

If the dust in NGC1068, IRAS08572 and IRAS19254 were formed by a
completely different mechanism from Galactic dust and/or underwent
different processing, then it is certainly possible that this would be
apparent from the profile of the 3.4$\mu$m feature.  Instead we
observe that the bands in these galaxies are as good a match to each
other and to Galactic spectra as are any laboratory spectra. Rather
than different processes which coincidentally produce materials with
almost identical features, the similarity of the feature in these
extragalactic lines of sight suggests similar dust formation and/or
processing mechanisms to those responsible for the feature in the
Galactic diffuse ISM. The mid-IR spectrum of NGC1068, which, like that
of the Galactic center, shows no indication of the deep absorption
features produced by many materials proposed as analogs of
carbonaceous dust, supports this interpretation.
The mismatch between a dust analog material, the Murchison meteorite
and NGC1068 emphasizes that, even though materials which bear a
close resemblance at 3.4$\mu$m can have very different mid-IR spectra,
gross differences are not observed between the Galactic diffuse ISM
and nuclear dust in NGC1068. This supports the suggestion that the
nuclear carbon-containing dust in NGC1068 is genuinely chemically
similar to that in the Galactic diffuse ISM. Further mid-IR studies of
other galaxies, notably IRAS08572, will be invaluable in determining
the full extent of this similarity.

\subsection{Dust production and survival in galaxies}
\label{discuss3}


That the dust in some Seyfert galaxies and ULIRGs appears to resemble
that towards the Galactic center and in CRL618 is intriguing. Can this
uniformity reveal anything about the specific production mechanism of
the hydrocarbon solids, the sensitivity of this mechanism to
environmental conditions, or the chemical and physical conditions of
the regions in which the carrier of the 3.4$\mu$m band is formed?

The debate over the origins of the carbonaceous fraction of
interstellar dust centers around two main formation mechanisms:
photochemical reactions in the icy grain mantles found in molecular
clouds, and condensation in stellar outflows followed by competing
processes of dehydrogenation and reformation of C\sbond H bonds in the
diffuse ISM \citep[e.g.][]{LG97,Ch98,Men03}. Laboratory experiments
show that irradiation of carbon-rich ice mixtures can indeed give rise
to material which absorbs at 3.4$\mu$m, but the profile of the band is
sensitive to the composition of the initial ice mixture. This
contrasts with the very uniform behavior of the astronomical
3.4$\mu$m band both in the galaxies examined here and in Galactic
lines of sight.  Laboratory analogs of dust produced in stellar
outflows involve processes such as exposing amorphous carbon grains or
graphite to hydrogen, or passing an electric or microwave-powered
discharge through gaseous methane or other hydrocarbon compound
\citep[see][and references therein, for examples]{PandA}. While there
is some variation in the 3.4$\mu$m band in the resulting materials, in
all cases but that of the CH$_{4}$ plasma/microwave discharge, this is
not very pronounced. However, in experiments mimicing dust formation
in carbon-rich circumstellar shells, \citet{Grishk} have shown that
the 3.4$\mu$m feature profile is influenced by the presence of
elements other than C and H. NH$_{3}$ and N$_{2}$ cause the individual
CH$_{2}$ and CH$_{3}$ subpeaks to become less distinct, whereas
O$_{2}$ causes absorption to be replaced by scattering-induced
emission.  The sensitivity of the 3.4$\mu$m band profile to
environmental conditions in both of these formation scenarios, in
contrast with the lack of variation in the feature now observed in a
number of galaxies, suggests that the conditions necessary for the
formation of the band carrier are common and quite invariant in
evolved stars and the diffuse ISM in a range of galaxies.

It is also possible that further processing of the dust grains after
their formation induces a degree of uniformity in their composition
and in the profile of the 3.4$\mu$m band. \citet{G95} present spectra
of ice residues exposed to the solar radiation field for four months
aboard the EURECA satellite and point out the very good resemblance
borne by the 3.4$\mu$m feature in these samples to that of the
Galactic diffuse ISM. The samples were synthesized from initial ice
mixtures of somewhat different compositions, and irradiated in the
laboratory before further irradiation in space. This raises the
possibility that prolonged UV exposure could cause initially varied
3.4$\mu$m bands to converge on the observed, astronomical
profile. However, in no case was the spectrum prior to solar
irradiation recorded, and the spectra of 13 other samples which also
flew were not presented, so this hypothesis currently remains
untested. As for dust produced in circumstellar shells, \citet{Grishk}
find that the differences in the 3.4$\mu$m band formed in the presence
of O and N often disappear when the dust is heated to
500$^{\circ}$C. If aliphatic hydrocarbons are synthesized in regions
of circumstellar envelopes which attain this temperature, this could
naturally explain the lack of variation in the resulting 3.4$\mu$m
feature. Observationally, the 3.4$\mu$m band has only been detected so
far in one PPNe \citep[CRL618;][]{LJ90,Ch98}; detection in a range of
evolved stars, with information on physical and chemical conditions,
could show whether the feature profile is insensitive to the
conditions under which it forms, or whether the necessary conditions
are in fact widespread.

As both the ice photolysis and stellar outflow routes to dust
production are capable of producing a 3.4$\mu$m band which matches the
observations quite well, that feature alone cannot be used to
distinguish decisively between these two mechanisms. However, as
discussed in \S\ref{MIR} and \S\ref{discuss.2}, the mid-IR spectrum of
NGC1068, like that of Sgr~A*, shows none of the strong mid-IR bands
which appear in spectra of photolysed ices. This means that none of
the ice residues currently proposed as analogs of carbonaceous dust
is really a viable candidate. The four months' irradiation of the
\citet{G95} samples is roughly equivalent to one passage of a dust
grain through the diffuse ISM before re-entering a molecular cloud,
and a typical grain will experience many such episodes during its
~10$^{9}$yr lifetime. \citet{G95} predict that further UV exposure
will remove O, N and H atoms, thus weakening the troublesome mid-IR
spectral features. Each time the grain enters a molecular cloud,
though, it will acquire a new ice coating containing precisely these
atoms, so it is not obvious that the mid-IR bands will be diminished
sufficiently to match the observations. The weight of evidence
concerning Galactic hydrocarbon-containing dust now tends to support
an origin in circumstellar environments and the diffuse ISM
\citep{Ch98,A99,PandA,Men03,Shenoy}, and the mid-IR spectrum of
NGC1068 supports this view.






That the 3.4$\mu$m absorption has been detected in ULIRGs and Seyfert
galaxies clearly means that the band carrier is capable of surviving
in these galaxies. ULIRGS are characterized by energy densities 3-4
orders of magnitude greater than those in the solar neighborhood
\citep{Maloney2}. Interestingly, \citet{Heck} find that certain
properties of the diffuse interstellar bands in starburst galaxies are
similar to those in Galactic lines of sight. Despite very high rates
of energy deposition into the ISM, the ratios of the strengths of the
bands and their correlation with reddening are roughly the same in the
starbursts as in the Galactic diffuse ISM. The authors speculate that
this is because, while the energy densities in these galaxies are
large, the ISM pressure is increased by a corresponding
amount. \citet{Maloney2} calculates that, like the energy densities,
the pressure in the ISM of ULIRGs is also 3-4 orders of magnitude
greater than the ISM pressure in the solar neighborhood, and that
this, coupled with a large shielding column of dust, can be sufficient
to protect even the fragile PAHs from X-ray destruction in AGN-powered
ULIRGs. In IRAS08572 and IRAS19254, where the 3.4$\mu$m band bears a
good resemblance to that in the Galactic diffuse ISM,
N$_{H}\ge10^{25}\rm{cm}^{-2}$ (Table~1), presumably providing a great
deal of shielding. 



As discussed in \S\ref{1068}, the 3.4$\mu$m feature in face-on Seyfert
2 galaxies is likely to arise in dust local to the active nucleus. The
exotic chemistry of the circumnuclear clouds in NGC1068 shows signs of
being influenced by the X-ray flux from the central engine
\citep{Stern,Usero}, so it is reasonable to ask whether X-ray emission
in AGN might also influence the hydrocarbon dust. Given that the
3.4$\mu$m feature is observed in absorption, however, the environment
in which the carbonaceous dust exists may not be as harsh as might be
expected. The nuclei of NGC1068 and NGC7674 are obscured at both hard
and soft X-ray energies (see Table~1), and H$_{2}$ emission is
concentrated on either side of the nuclear IR flux peak in NGC1068
\citep{Galla,Grata}. These observations suggest that the absorbing
dust is exposed to neither X-rays nor UV photons, which would be
attenuated by the large ($N_{H}>10^{25}{\rm cm}^{-2}$) column of
warmer dust between it and the nucleus. Similarly, the absence of
3.3$\mu$m PAH emission in the smallest-aperture L-band spectra of
NGC1068 could reflect not destruction of the molecules in a harsh
environment, but simply lack of an excitation mechanism. In some of
the other Seyfert galaxies the shielding column is considerably
smaller than in NGC1068 and NGC7674.  These galaxies are closer to
edge-on, meaning that some of the dust in the line of sight may be far
from the nucleus. As the scale heights of dust in spiral galaxies tend
to be small \citep{Alton}, though, these spectra are likely to still
contain a substantial contribution from nuclear dust. This suggests
that the hydrocarbons in the dust in these galaxies are robust enough
to survive unchanged in quite harsh conditions.

\section{Conclusions}
\label{conc}

We have shown that the 3.4$\mu$m C\sbond H bond stretch of
interstellar dust in the Seyfert galaxy, NGC1068, and two ULIRGs,
IRAS08572 and IRAS19254, has a very similar profile to that seen in
Galactic carbonaceous material. Of the other galaxies in which the
3.4$\mu$m feature has been detected at much lower S/N, several also
have feature profiles consistent with that of NGC1068. Although minor
differences are seen between IRAS19254 and other objects, these
differences are small compared with the variations that are capable of
occurring in hydrocarbon chains of varying lengths and in differing
chemical environments.  This implies that the hydrocarbon portion of
the carbonaceous dust in these galaxies bears a close resemblance to
that which exists in our own Galaxy, probably as a result of the dust
having a similar formation and processing history to that in the
Galactic diffuse ISM. Examination of the mid-IR spectrum of NGC1068
reveals none of the strong absorption bands characteristic of many
materials proposed as analogs of carbonaceous dust. Although features
like those present in the mid-IR spectrum of the Galactic center would
be undetectable in the available data for NGC1068, the lack of major
differences is consistent with the suggestion that the dust in this
AGN is chemically very similar to dust in the Galactic diffuse ISM. It
also suggests that condensation in circumstellar environments, rather
than processing of icy grain mantles, is the dominant source of
carbonaceous dust in many galaxies. Higher S/N spectroscopy at 3.4 and
5-9$\mu$m, particularly of the galaxies whose lower S/N spectra hint
at different feature profiles, will prove invaluable in assessing
whether this remarkable chemical similarity is a common feature of
active, ultraluminous and quiescent galaxies alike.

\acknowledgments
We thank the anonymous referee for a prompt and constructive report.
RM thanks PPARC for a PhD studentship. YP gratefully acknowledges
support for this work from NASA's Exobiology program (344-38-12-09).




\clearpage

\objectname{GC~IRS6}
\objectname{CRL618}
\objectname{IRAS04385-0828}
\objectname{IRAS05189-2524}
\objectname{IRAS08572+3915}
\objectname{IRAS19254-7245}
\objectname{NGC1068}
\objectname{NGC3094}
\objectname{NGC5506}
\objectname{NGC7172}
\objectname{NGC7674}
\objectname{NGC7479}
\objectname{UGC5101}
\objectname{Mrk463}


\begin{thebibliography}{}


\bibitem[Adamson, Whittet \& Duley (1990)]{A90}Adamson, A. J.,
Whittet, D. C. B., \& Duley, W. W., 1990, MNRAS, 243, 400
\bibitem[Adamson et al. (1999)]{A99}Adamson, A. J., Whittet, D. C. B.,
Chrysostomou, A., Hough, J. H., Aitken, D. K., Wright, G. S., \&
Roche, P. F. 1999, ApJ, 512, 224
\bibitem[Allamandola et al. (1988)Allamandola, Sandford \& Valero]{A88}Allamandola,
L. J., Sandford, S. A., \& Valero, G. J. 1988, Icar, 76, 225
\bibitem[Alpert, Keiser \& Szymanski (1970)]{Alp} Alpert, N., Keiser,
W., \& Szymanski, H. 1970, Theory and Practice of Infrared Spectroscopy
(2nd Ed.; New York: Plenum Press)
\bibitem[Alton et al. (2000)]{Alton}Alton, P. B., Rand, R. J.,
Xilouris, E. M., Bevan, S., Ferguson, A. M., Davies, J. I., \& Bianchi,
S. 2000, A\&AS, 145, 83
\bibitem[Antonucci \& Miller (1985)]{AM85}Antonucci, R, \& Miller, J.,
1985, ApJ, 297, 621
\bibitem[Bassani et al. (1999)]{B99}Bassani, L., Dadina, M., Maiolino,
R., Salvati, M., Risaliti, G., della Ceca, R., Matt, G., \& Zamorani, G.,
1999, ApJS, 121, 473
\bibitem[Bridger, Wright \& Geballe (1994)]{Bridger}Bridger, A.,
Wright, G. S., \&  Geballe, T. R. 1994, in ASSL Vol. 190, Infrared
Astronomy with Arrays, The Next Generation, ed. Ian S. McLean
(Dordrecht: Kluwer Academic Publishers), p.537
\bibitem[Butchart et al. (1986)]{Butch} Butchart, I., McFadzean,
A. D., Whittet, D. C. B., Geballe, T. R., \& Greenberg, J. M. 1986, A\&A,
154L, 5
\bibitem[Cernicharo et al. (2001)]{Cerni} Cernicharo, J., Heras, A.
M., Pardo, J. R., Tielens, A. G. G. M., Guélin, M., Dartois, E.,
Neri, R., \& Waters, L. B. F. M. 2001, ApJ, 546L, 127
\bibitem[Charmandaris et al. (2002)]{Charm}Charmandaris, V., Laurent,
O., Le Floc'h, E., Mirabel, I. F., Sauvage, M., Madden, S. C.,
Gallais, P., Vigroux, L., \& Cesarsky, C. J. 2002, A\&A, 391, 429
\bibitem[Chiar et al. (1998)]{Ch98}Chiar, J. E., Pendleton, Y. J.,
Geballe, T. R., \& Tielens, A. G. G. M. 1998, ApJ, 507, 281
\bibitem[Chiar et al. (2000)]{Chiar00}Chiar, J. E., Tielens,
A. G. G. M., Whittet, D. C. B., Schutte, W. A., Boogert, A. C. A.,
Lutz, D., van Dishoeck, E. F., \& Bernstein, M. P. 2000, ApJ, 537, 749
\bibitem[Chiar et al. (2002)]{Chiar02}Chiar, J. E., Adamson, A. J.,
Pendleton, Y. J., Whittet, D. C. B., Caldwell, D. A., \& Gibb, E. L.,
2002, ApJ, 570, 198
\bibitem[Cronin \& Pizzarello (1990)]{Murch1}Cronin, J. R.,
\& Pizzarello, S. 1990, GeCoA, 54, 2859
\bibitem[de Vaucouleurs et al (1991)]{RC3}de Vaucouleurs, G., de
Vaucouleurs, A., Corwin, H., Buta, R., Paturel, G., \& Fouque, P. 1991,
Third Reference Catalogue of Bright Galaxies (Berlin: Springer) 
\bibitem[de Vries et al. (1993)]{Murch}de Vries, M. S., Reihs, K.,
Wendt, H. R., Golden, W. G., Hunziker, H. E., Fleming, R., Peterson,
E., \& Chang, S. 1993, GeCoA, 57, 933
\bibitem[Dartois et al. (2004)]{Dartois} Dartois, E., Marco, O.,
Mu\~{n}oz-Caro, G., Brooks, K., Deboffle, D., \& d'Hendecourt, L.
2004, A\&A, in press
\bibitem[Devillard (1997)]{Dev} Devillard, N. 1997,  The Messenger, 87 
\bibitem[Furton, Laiho \& Witt (1999)]{Furton}Furton, D. G.,
Laiho, J. W., \& Witt, A. N. 1999, ApJ, 526, 752
\bibitem[Galliano \& Alloin (2002)]{Galla}Galliano, E., \& Alloin, D.,
2002, A\&A, 393, 43
\bibitem[Garc\'{i}a-G\'{o}mez, Athanassoula \&
Barber\`{a} (2002)]{Garcia}Garc\'{i}a-G\'{o}mez, C., Athanassoula, E., \&
Barber\`{a}, C. 2002, A\&A, 389, 68
\bibitem[Gratadour et al. (2003)]{Grata}Gratadour, D., Cl\'{e}net, Y.,
Rouan, D., Lai, O., \& Forveille, T. 2003, A\&A, 411, 335
\bibitem[Greenberg et al. (1995)]{G95}Greenberg, J. M., Li, A.,
Mendoza-Gomez, C. X., Schutte, W. A., Gerakines, P. A., de
\& Groot, M. 1995, ApJ, 455L, 177
\bibitem[Grishko \& Duley (2002)]{Grishk}Grishko, V. I., \& Duley, W. W.,
2002, ApJ, 568, 448
\bibitem[Heckman \& Lehnert (2000)]{Heck}Heckman, T., \& Lehnert, M.,
2000, ApJ, 557, 690
\bibitem[Imanishi et al. (1997)]{Im97}Imanishi, M., Terada, H.,
Sugiyama, K., Motohara, K.,Goto, M., \& Maihara, T. 1997, PASJ, 49, 69
\bibitem[Imanishi et al. (1996)]{Im96}Imanishi, M., Sasaki, Y., Goto,
M., Kobayashi, N., Nagata, T., \& Jones, T., 1996, AJ, 112, 235
\bibitem[Imanishi (2000a)]{Im00a}Imanishi, M. 2000, MNRAS, 313, 165
\bibitem[Imanishi (2000b)]{Im00b}Imanishi, M. 2000, MNRAS, 319, 331
\bibitem[Imanishi \& Dudley (2000)]{ID00}Imanishi, M., \& Dudley, C.,
2000, ApJ, 545, 701
\bibitem[Imanishi et al. (2001)Imanishi, Dudley \& Maloney]{Im01}Imanishi, M.,
Dudley, C., \& Maloney, P. 2001, \apj, 558L, 93
\bibitem[Imanishi (2002)]{Im02}Imanishi, M. 2002, \apj, 569, 44
\bibitem[Imanishi (2003)]{Im032}Imanishi, M. 2003, ApJ, 599, 918
\bibitem[Imanishi et al. (2003)]{Imetal}Imanishi, M., Terashima, Y.,
Anabuki, N., \& Nakagawa, T. 2003, ApJ, 596L, 167
\bibitem[Ishii et al. (2002)]{Ishii}Ishii, M., Nagata, T.,
Chrysostomou, A., \& Hough, J. H. 2002, AJ, 124, 2790
\bibitem[Kinney et al. (2000)]{Kinney}Kinney, A. L., Schmitt, H. R.,
Clarke, C. J., Pringle, J. E., Ulvestad, J. S., \& Antonucci, R. R. J.,
2000, ApJ, 537, 152
\bibitem[Le Floc'h et al. (2001)]{LeF}Le Floc'h, E., Mirabel, I. F.,
Laurent, O., Charmandaris, V., Gallais, P., Sauvage, M., Vigroux, L., \&
Cesarsky, C. 2001,A\&A, 367, 487
\bibitem[Lequeux \& Jourdain de Muizon (1990)]{LJ90}Lequeux, J., \&
Jourdain de Muizon, M. 1990, A\&A, 240L, 19
\bibitem[Li \& Greenberg (1997)]{LG97} Li, A., \& Greenberg, J. M. 1997,
A\&A, 323, 566
\bibitem[Lutz et al. (2000)]{Lutz}Lutz, D., Sturm, E., Genzel, R.,
Moorwood, A. F. M., Alexander, T., Netzer, H., \& Sternberg, A. 2000,
ApJ, 536, 697
\bibitem[Maloney (1999)]{Maloney2}Maloney, Philip R. 1999, Ap\&SS,
266, 207
\bibitem[Marco \& Brooks (2003)]{MB03}Marco, O., \& Brooks, K.,
2003, A\&A 398, 101
\bibitem[McFadzean et al. (1989)]{McFadz}McFadzean, A. D., Whittet,
D. C. B., Bode, M. F., Adamson, A. J., \& Longmore, A. J. 1989, MNRAS,
241, 873
\bibitem[Mennella et al. (1999)]{Men99}Mennella, V., Brucato, J. R.,
Colangeli, L., \& Palumbo, P. 1999, ApJ, 524L, 71
\bibitem[Mennella et al. (2003)]{Men03}Mennella, V., Baratta, G. A.,
Esposito, A., Ferini, G., \& Pendleton, Y. J. 2003, ApJ, 587, 727
\bibitem[Mizutani, Suto \& Maihara (1994)]{Mizutani}Mizutani, K.,
Suto, H., \& Maihara, T 1994, ApJ, 421, 475
\bibitem[Nagar \& Wilson (1999)]{Nagar} Nagar, N., \& Wilson, A. 1999,
ApJ, 516, 97
\bibitem[Pendleton et al. (1994)]{P94}Pendleton, Y. J., Sandford,
S. A., Allamandola, L. J., Tielens, A. G. G. M., \& Sellgren, K. 1994,
ApJ, 437, 683
\bibitem[Pendleton (1996)]{P96}Pendleton, Y. J. 1996, in ASSL
Vol. 209, New Extragalactic Perspectives in the New South Africa,
ed. D. Block \& J. Greenberg (Dordrecht: Kluwer Academic Publishers),
p.135
\bibitem[Pendleton (1997)]{P97}Pendleton, Y. J. 1997, OLEB, 27, 53
\bibitem[Pendleton \& Allamandola (2002)]{PandA} Pendleton, Y. J.,
\& Allamandola, L. J. 2002, ApJS, 138, 75
\bibitem[Rawlings, Adamson \& Whittet~(2003)]{R03}Rawlings, M. G.,
Adamson, A. J., \& Whittet, D. C. B. 2003, MNRAS, 341, 1121
\bibitem[Risaliti et al.~(2003)]{Ris}Risaliti, G., Maioliono, R.,
Marconi, A., Bassani, L., Berta, S., Braito, V., Della Ceca, R.,
Franceschini, A., \& Salvati, M. 2003, ApJ, 595L, 17
\bibitem[Risaliti et al. (2000)]{Ris00}Risaliti, G., Gilli, R.,
Maiolino, R., \& Salvati, M. 2000, A\&A, 357, 13
\bibitem[Rouan et al. (2004)]{Rouan2}Rouan, D., Lacombe, F., Gendron,
E., Gratadour, D., Clénet, Y., Lagrange, A.-M., Mouillet, D., Boisson,
C., Rousset, G., Fusco, T., and 7 coauthors 2004, A\&A, 417, 1
\bibitem[Sakata et al. (1987)]{S87}Sakata, A., Wada, S., Onaka, T., \&
Tokunaga, A. 1987, ApJ, 320L, 63
\bibitem[Sandford et al. (1991)]{S91}Sandford, S. A., Allamandola,
L. J., Tielens, A. G. G. M., Sellgren, K., Tapia, M., \& Pendleton, Y.,
1991, ApJ, 371, 607
\bibitem[Sandford, Pendleton \& Allamandola (1995)]{S95}Sandford,
S. A., Pendleton, Y. J., \& Allamandola, L. J. 1995, ApJ, 440,
697
\bibitem[Scoville et al. (2000)]{Scov}Scoville, N. Z., Evans, A. S.,
Thompson, R., Rieke, M., Hines, D. C., Low, F. J., Dinshaw, N.,
Surace, J. A., \& Armus, L. 2000, AJ, 119, 991
\bibitem[Shenoy et al. (2003)]{Shenoy}Shenoy, S. S., Whittet,
D. C. B., Chiar, J. E., Adamson, A. J., Roberge, W. G., \& Hassel, G. E.,
2003, ApJ, 591, 962
\bibitem[Sternberg, Genzel \& Tacconi (1994)]{Stern} Sternberg, A.,
Genzel, R., \& Tacconi, L. 1994, ApJ, 436L, 131
\bibitem[Sturm et al. (2000)]{Sturm}Sturm, E., Lutz, D., Tran, D.,
Feuchtgruber, H., Genzel, R., Kunze, D., Moorwood, A. F. M., \& Thornley,
M. D. 2000, A\&A, 358, 481
\bibitem[Tomono et al. (2001)]{Tom}Tomono, D., Doi, Y.,
Usuda, T., \& Nishimura, T. 2001, ApJ, 557, 637
\bibitem[Usero et al. (2003)]{Usero} A. Usero, S. Garcia-Burillo,
A. Fuente, \& J. Martin-Pintado 2003, astro-ph/0402556
\bibitem[Whittet et al. (1997)]{Wh97}Whittet, D. C. B., Boogert,
A. C. A., Gerakines, P. A., Schutte, W., Tielens, A. G. G. M., de
Graauw, Th., Prusti, T., van Dishoeck, E. F., Wesselius, P. R., \&
Wright, C. M. 1997, ApJ, 490, 729 
\bibitem[Whittle (1992)]{Whittle} Whittle, M. 1992, ApJS, 79, 49
\bibitem[Wright et al. (1996)]{boss}Wright, G., Bridger, A., Geballe,
T., \& Pendleton, Y. 1996, in ASSL Vol. 209, New Extragalactic
Perspectives in the New South Africa, ed. D. Block \& J. Greenberg
(Dordrecht: Kluwer Academic Publishers), p.143
\end{thebibliography}
\end{document}